\documentclass[twocolumn]{aastex62}

\submitjournal{ApJ}
\shorttitle{The Origins of Gas Accreted by SMBH}
\shortauthors{Choi et al.}

\def\galaxiespprox{\mathrel{\vcenter{\offinterlineskip \hbox{$>$}
    \kern 0.3ex \hbox{$\sim$}}}}
\def\lapprox{\mathrel{\vcenter{\offinterlineskip \hbox{$<$}
    \kern 0.3ex \hbox{$\sim$}}}}
\newcommand{\beq}{\begin{equation}} 
\newcommand{\eeq}{\end{equation}}

\def\eps{\epsilon}

\def\Sig15{\Sigma_{1.5}}

\def\kpch{{\rm\thinspace kpc} \thinspace \it h^{\rm -1}}
\def\Msun{\hbox{$\thinspace \rm M_{\odot}$}}
\def\Msunh{\hbox{$\thinspace \rm \thinspace M_{\odot}\thinspace \it h^{\rm -1}$}}

\def\rvir{r_{\rm vir}}

\def\strom{Str$\rm \ddot{o}$mgren$\thinspace$}

\def\delm12{M_{12}}
\def\sigm1{\sigma(m_{1})}

\begin{document}

\title{The Origins of Gas Accreted by Supermassive Black Holes: the Importance of Recycled Gas}

\correspondingauthor{Ena Choi}
\email{enachoi@uos.ac.kr}
\author[0000-0002-8131-6378]{Ena Choi}
\affiliation{Department of Physics, University of Seoul, 163 Seoulsiripdaero, Dongdaemun-gu, Seoul 02504, Republic of Korea}

\author{Rachel S. Somerville}
\affil{Center for Computational Astrophysics, Flatiron Institute, 162 5th Ave., 
    New York, NY 10010, USA}
    
\author{Jeremiah P. Ostriker}
\affil{Department of Astronomy, Columbia University, 
    550 West 120th Street, 
    New York, NY 10027, USA}
\affil{Department of Astrophysical Sciences, Princeton University, 
    Princeton, NJ 08544, USA}     

\author{Michaela Hirschmann}
\affil{Institute for Physics, Laboratory for Galaxy Evolution and Spectral modelling, Ecole Polytechnique Federale de Lausanne,\\
Observatoire de Sauverny, Chemin Pegasi 51, 1290 Versoix, Switzerland}
\affil{INAF - Osservatorio Astronomico di Trieste, via G. B. Tiepolo 11, I-34143 Trieste, Italy}

\author{Thorsten Naab}
\affil{Max-Planck-Institut f\"ur Astrophysik,
    Karl-Schwarzschild-Strasse 1, 85741 Garching, Germany}

\begin{abstract}
We investigate the fueling mechanisms of supermassive black holes (SMBHs) by analyzing ten zoom-in cosmological simulations of massive galaxies, with stellar masses $10^{11-12} M_{\odot}$ and SMBH masses $10^{8.9-9.7}$ at $z=0$ and featuring various major and minor merger events. By tracing the gas history in these simulations, we categorize the gas accreted by the central SMBHs based on its origin. Gas that belonged to a different galaxy before accretion onto the BH is labeled as (1) ``external," while smoothly accreted cosmic gas is classified as (2) ``smooth." Gas produced within the primary halo through stellar evolution and subsequently accreted by the SMBH is classified as (3) ``recycled." Our analysis, which included stellar feedback, reveals that the primary fuel source for SMBHs is the recycled gas from dying stars. This recycled gas from stars in the inner region of the galaxy readily collapses toward the center, triggering starbursts, and simultaneously fueling the SMBH. Galaxy mergers also play a crucial role in fueling SMBHs in massive galaxies as SMBHs in massive halos tend to accrete a higher fraction of external gas from mergers compared to smoothly accreted gas. However, on average, it takes approximately 1.85 Gyr for external gas to enter the main galaxy and accrete onto the SMBH. Considering the presence of various other gas triggers for AGN activity alongside this time delay, the association between AGN and mergers may not always be obvious.
\end{abstract}
\section{Introduction} \label{sec:intro}

The strong connections between the cosmic evolution of supermassive black holes (SMBHs) and their host galaxies have been investigated for  more than two decades. First of all, observational evidence has shown that almost every large galaxy has a SMBH at its center \citep[e.g.][]{Kormendy1995,1998AJ....115.2285M}.  The SMBH mass is strongly correlated with the spheroidal mass, velocity dispersion $\sigma$ and other properties of the galaxy \citep[e.g.][]{Gebhardt2000,Haring2004,2013ARA&amp;A..51..511K,Woo2013}. Furthermore, the global cosmic star formation history and cosmic black hole accretion history follow each other remarkably closely, rising from the epoch of reionization to $z \sim 2-3$, then steeply declining to the current epoch, suggesting that SMBH and galaxies may have co-evolved.

Energetic feedback from accreting black holes, commonly called Active Galactic Nuclei (AGN) feedback, has become a key ingredient in galaxy evolution simulations \citep{2005ApJ...620L..79S,sijacki2007,somerville2008a}. AGN feedback prevents over-cooling and excessive star formation in galaxies and also self-regulates the black hole growth \citep[e.g.][]{Booth2009,Gaspari2011,Choi2012a,Dubois2012,Appleby2020}. Many contemporary large-volume simulations such as EAGLE, TNG, SIMBA, HorizonAGN and Magneticum all include the growth of black holes and feedback from them, and were able to reproduce these gigantic $10^{8-9}$ $\Msun$ supermassive blackholes at $z=0$ \citep{Hirschmann2014,Schaye2015a,2016MNRAS.463.3948D,Weinberger2018,Dave2019}. For a more detailed review of black hole growth in cosmological simulations, see \citet{Somerville2015,2017ARA&amp;A..55...59N}.

How do SMBHs grow? The origin of SMBHs remains uncertain, but they are thought to grow primarily through two mechanisms: accretion and mergers. Black holes grow by accreting matter from the surrounding gas, which spirals toward the black hole and converts gravitational potential energy into heat, providing the principal source of radiation for active galactic nuclei (AGNs). They also grow by merging with other black holes when two galaxies merge, bringing their respective black holes together. AGN pairs have been observed in X-ray, radio, and optical imaging and spectroscopy \citep[e.g.][]{Komossa2003}, and more than 100 stellar mass black hole mergers have been discovered since the Laser Interferometer Gravitational-Wave Observatory (LIGO). Based on the Soltan argument \citep{1982MNRAS.200..115S}, it is expected that mass accretion plays a substantial role in black hole growth. Moreover, the dominance of mass accretion as the primary source of black hole mass is necessary to explain the evolution of the X-ray luminosity function of AGN \citep[e.g.][]{Hirschmann2014}. However, alternative studies suggest that the correlations between SMBH mass and other galaxy properties could be set by mergers \citep{Hirschmann2010,Jahnke2011}.

However, it is still unclear how SMBH of more than nearly a billion solar masses were formed, both from theoretical and observational perspectives. The contribution of black hole mergers and gas accretion remains uncertain, and a comprehensive understanding of the origin of the gas that fuels supermassive black holes (SMBHs) is lacking. It remains unclear whether the primary source of gas for SMBH formation is externally supplied through galaxy mergers or gradually accreted from intergalactic medium (IGM) filaments. To shed light on this matter, this study aims to use cosmological hydrodynamic simulations of massive galaxies and SMBH formation.

The simulation set utilized in this study comprises 10 simulations of the formation of massive galaxies with stellar masses ranging from $1-5 \times 10^{11}$ and containing SMBHs of approximately $10^{8.9-9.7}$ solar masses at $z=0$. The formation processes of the black holes are examined in detail from $z=5$ to $z=0$ to investigate the source of the gas that fuels the black holes. Specifically, we track individual gas particles that feed the black holes and classify their origins into four categories: external gas that enters the host galaxy through galaxy mergers, diffuse gas that is smoothly accreted into the host galaxy, recycled gas produced within the galaxy by stellar evolution, and early accretion of gas deposited in the main halo before $z=5$.

The methodology of particle tracing for black hole accretion analysis has already been employed in a study by \cite{Bellovary2013}. In their analysis, \cite{Bellovary2013} investigated the formation process of three SMBHs with masses of $10^{6-7} \thinspace \Msun$, simulated up to $z = 4$, using particle tracing. They found that intermediate-mass black holes at high redshifts grew primarily through the accretion of cold flows directly entering the galaxy from the IGM. More recently, \cite{Sanchez2018} used the same particle tracing method to study the formation of a $10^7 \thinspace \Msun$ SMBH at $z = 0$ in a Milky Way-sized galaxy simulation. Their work demonstrated that black holes within galaxies with a merger-rich history, including two major mergers, are primarily formed by accreted gas through the merger. However, this study only analyzed a single simulation, making it impossible to obtain information about other galaxies with different formation histories and the origins of their black holes.

Therefore, the purpose of this study is to investigate the formation origins of multiple SMBHs, each located at the center of simulated galaxies, with a particular focus on high-mass SMBHs of the order of $10^9 \thinspace \Msun$ in massive galaxies. Our analysis includes a novel examination of recycled gas in a statistical sample of high resolution simulations, which has not been explored in previous studies. In elliptical galaxies with an abundance of old stellar populations, the effects of SN and AGB winds from evolved stars are inevitable. As discussed by \cite{Ciotti2007}, recycled gas resulting from this stellar evolution can trigger late-time AGN flares and star formation. We conduct a detailed tracking of this recycled gas on a galaxy by galaxy basis to determine its impact on central black hole growth.

In summary, this paper aims to investigate the gas accretion origins of SMBHs by closely examining what they {\it eat} for their meal in cosmological hydrodynamic simulations of galaxy formation. We trace the history of the accreted gas and categorize them into four origins: external, recycled, smooth, and early, and examine the importance of each. Section~\ref{sec:method} discusses the simulation and particle tracing methods used in this study in detail. In Section ~\ref{sec:results}, we present the results of our particle tracing analysis. Section~\ref{sec:discussion} includes a discussion of our findings, and finally, in Section~\ref{sec:summary}, we conclude with a summary.

\section{Simulation and Method}\label{sec:method}
\subsection{Simulation}\label{subsec:simulation}
We use a subset of the suite of cosmological hydrodynamical simulations of massive galaxy and supermassive black hole formation presented in \citet{Choi2017}. In this section, we provide an overview of the simulations, focusing on the black hole formation and growth. We refer the reader to \citet{Choi2017} for more details on the simulation suite. 

The initial conditions of zoom-in regions of massive halos are adopted from \citet{Oser2010}. The halos of interest are selected from the dark matter only simulation with a comoving periodic box size of $L=100$~Mpc, simulated with cosmological parameters of WMAP3 \citep[][$h=0.72, \;\Omega_{\mathrm{b}}=0.044, \; \Omega_{\mathrm{dm}}=0.26, \;\Omega_{\Lambda}=0.74, \; \sigma_8=0.77 $, and $\mathrm{n_s}=0.95$]{2007ApJS..170..377S}. The high-resolution zoom-in initial conditions are constructed by tracing back the evolution of halos of interest with time, and replacing all particles within $2 \times \rvir$ of the halos with high resolution dark matter and baryonic particles. Therefore, the refined high resolution zoom-in initial conditions fully include the cosmological assembly of the halo of interest. 

The refined initial conditions are simulated with SPHGal \citep{2014MNRAS.443.1173H}, a modified version of the parallel smoothed particle hydrodynamics (SPH) code GADGET-3 \citep{2005MNRAS.364.1105S}. To overcome the numerical fluid-mixing problems of classical SPH codes, this code incorporates a number of improvements: a density-independent pressure-entropy SPH formulation from \citep{2001MNRAS.323..743R}, an improved artificial viscosity \citep{2010MNRAS.408..669C}, and an artificial thermal conductivity \citep{2012MNRAS.422.3037R}. The mass resolutions are $m_{\mathrm{dm}} = 2.5 \times 10^{7} \Msunh$ for dark matter particles, and $m_{*,gas}=4.2 \times 10^{6} \Msunh$ for gas and star particles in the simulation suite. The co-moving gravitational softening lengths are $\eps_{\mathrm{halo}} = 0.89 \rm \kpch$ for the dark matter and $\eps_{\mathrm{gas,star}} = 0.4 \rm \kpch $ for the gas and star particles.

The star formation and chemical enrichment model are adopted from \cite{2013MNRAS.434.3142A}. The chemical enrichment is explicitly calculated for 11 different species, H, He, C, N, O, Ne, Mg, Si, S, Ca, and Fe in star and gas particles. The output masses from Type~I Supernovae (SNe), Type~II SNe, and asymptotic giant branch (AGB) stars gradually enrich the surrounding gas, with the chemical yields adopted from \citet{1999ApJS..125..439I,1995ApJS..101..181W,2010MNRAS.403.1413K} respectively. We note that with our assumed IMF, approximately 30 \% of the total mass in newly formed stars will eventually be released back into the gas phase through stellar winds over $\sim$13 Gyr of stellar evolution.

The recycled stellar ejecta from stellar particles is returned to the surrounding gas particles, resulting in an increase in the mass of the gas particles with enriched metal content. Consequently, the mass of the stellar particles becomes smaller than the initial baryon particle resolution, while the gas particles tend to have larger masses. Occasionally, a subset of these gas particles may accumulate enough stellar ejecta to reach twice the initial resolution mass. In such cases, the gas particle is divided into two equal-mass gas particles with identical physical properties to accommodate the continuous addition of stellar ejecta.

The simulations also include stellar kinetic feedback using the approach presented in  \citet{2017ApJ...836..204N}. This model incorporates the feedback from four elements, i) the winds from young massive stars, ii) UV heating within \strom spheres of young stars, iii) three-phase Supernova remnant input from both type I and type II SN feedback, and iv) outflows from low-mass AGB stars. Metal diffusion is included in the simulation following \cite{2013MNRAS.434.3142A}, i.e., the metal-enriched gas particles can mix their metals with neighboring gas via turbulent diffusion.

At the center of newly emerging dark matter halos with mass above $1\times10^{11} \Msunh$, new black hole particles with a mass of $10^5 \Msunh$ are seeded. Black holes are allowed to grow via two channels: mergers with other black holes and gas accretion. The simulation permits black hole-black hole mergers only if the two black hole particles are within their respective SPH smoothing lengths and possess relative velocities that are lower than the local sound speed. A Bondi-Hoyle-Lyttleton parameterization \citep{1939PCPS...34..405H,1944MNRAS.104..273B,1952MNRAS.112..195B} is adopted to estimate the gas accretion rate onto a black hole. In addition, the model includes a soft Bondi criterion that statistically limits the accretion to the gas within the Bondi radius \citep{Choi2012a}. We calculate the Bondi radius of each black hole, account for the size of the gas particle with its smoothing length, and allow the full accretion only when the total volume of a gas particle is included within the Bondi radius. If a gas particle volume is only partially included within the Bondi radius, the accretion probability of that gas particle is reduced. If a gas particle volume is not within the Bondi radius, the gas particle is not accreted. Additionally, we include the free-fall timescale in the accretion rate to account for the time it takes for the particle to be accreted onto the black hole.

The accretion rate is not artificially limited to the Eddington rate, but instead the simulation includes the Eddington force that pushes (notional) electrons in the gas particles radially away from the black hole. Therefore, the black hole occasionally accretes gas at a Super-Eddington rate, but the feedback processes summarized below quickly regulate the gas accretion.

The simulations incorporate mechanical AGN feedback, which imparts mass and momentum to the surrounding gas \citep[see][for details.]{Choi2012a,Choi2014}. Mechanical AGN feedback is very effective at quenching SF and self-regulating BH growth \citep{Ostriker2010a,Choi2015a}. This model mimics the observed strong outflows \cite[e.g.][]{Arav2020} launched by radiation from radiatively efficient accretion onto a black hole \citep[e.g.][]{2000ApJ...543..686P,2004ApJ...616..688P}. We also employ a time-step limiter, which requires that all neighboring particles have a similar time step. This ensures that ambient particles do not remain inactive when a shock, such as a strong wind driven by AGN, travels through. 

The simulations also include Compton and photoionization heating and the associated radiation pressure effect of moderately hard X-ray radiation ($\sim 50$~keV) from the accreting black hole following \citet{2004MNRAS.347..144S,2005MNRAS.358..168S}. This mechanical and radiative AGN feedback model has been shown to effectively suppress star formation in cosmological simulations of massive galaxies \citep{Choi2015a,Choi2017}.

These galaxy simulations are calibrated to align with the local $M_{\rm BH} - \sigma$ relation and the stellar-to-halo mass relation. For an in-depth examination of the physical properties of the complete galaxy sample, please refer to \citet{Choi2017}.

We identify haloes and subhalos in our simulations using the ROCKSTAR Halo Finder \citep{2013ApJ...762..109B}. It utilizes adaptive hierarchical refinement a friends-of-friends algorithm in six phase-space dimensions and allows for a better tracking of substructure.

\begin{figure}
\plotone{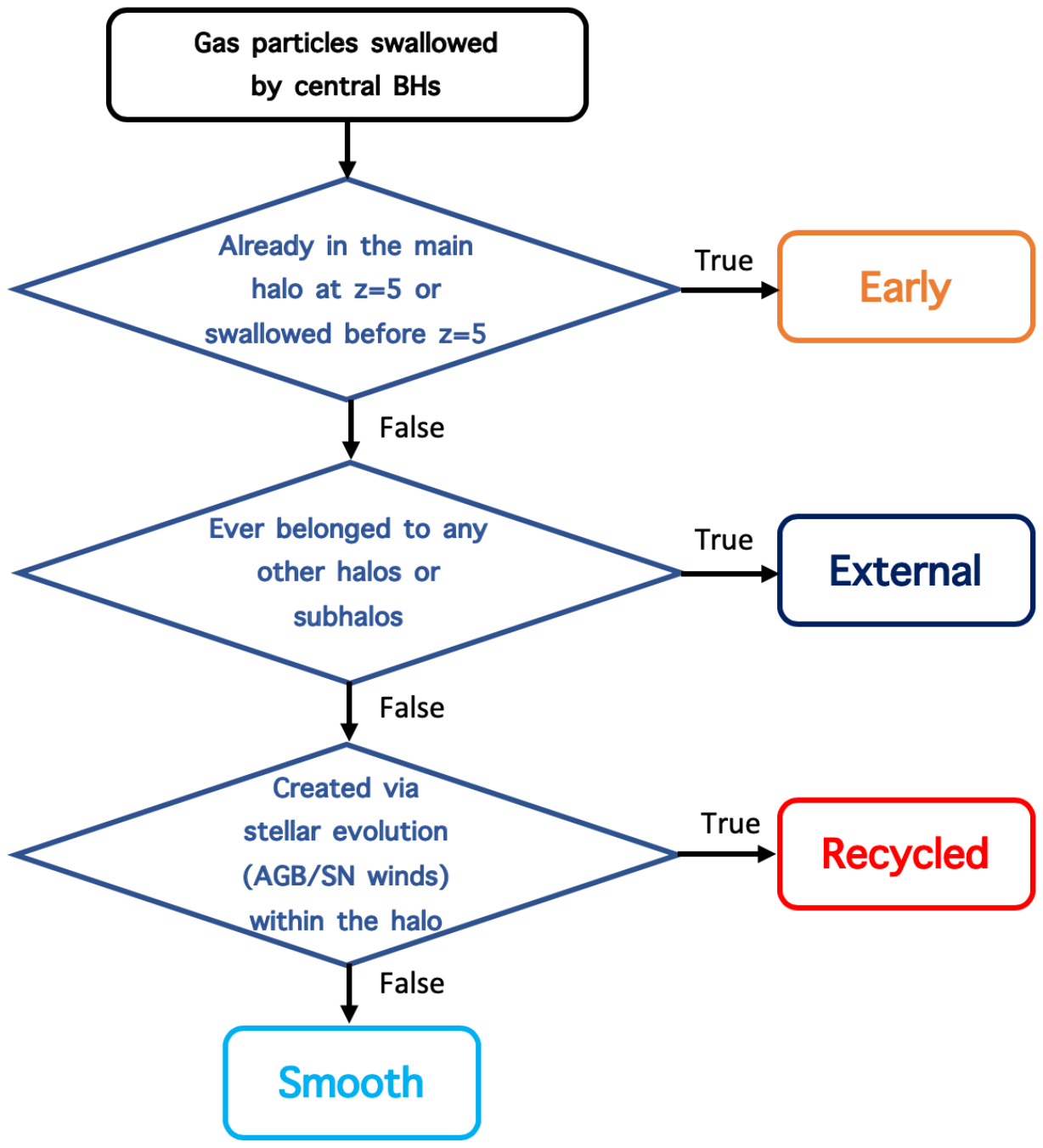}
\caption{Flowchart of the gas origin classification method. We classify the origin of gas accreted by black holes into four categories: early, external, recycled, and smooth. \label{fig:flowchart}}
\end{figure}

\begin{figure*}
\gridline{\fig{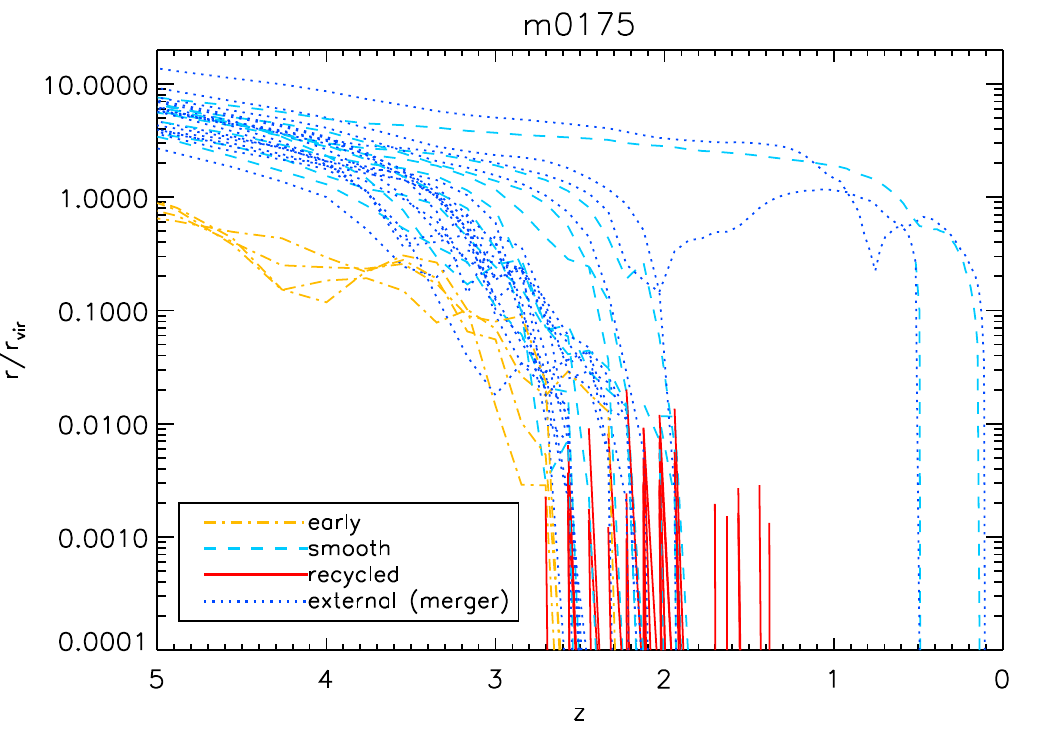}{0.5\textwidth}{}
     \fig{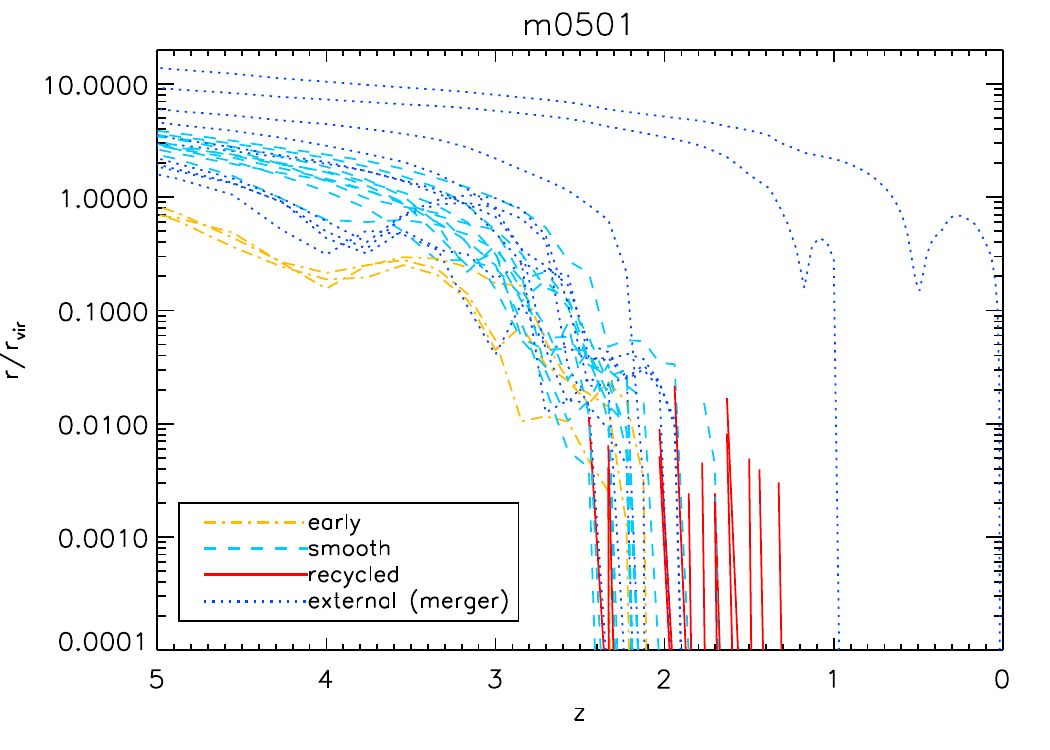}{0.5\textwidth}{}}
\caption{Trajectories of individual gas particles over cosmic time before they accrete to the primary black hole in two example galaxies, m0175, and m0501. The y-axis shows the distance of the gas particle from the central black hole in units of the virial radius of the primary halo. The accreted gas originating from recycled gas, external gas, smoothly accreted gas, and early accretion are shown in red, blue, light blue, and orange, respectively. For clarity, we only show a few gas particles for each category as an example. Many external gas particles orbit around the main halo before they accrete to the primary black hole, which sits at the bottom of this plot. All recycled gas is born within the galaxy, $r_{10}=0.1 \times \rvir$ of the primary halo by definition, then almost immediately accretes to the black hole. \label{fig:trajectory}} %
\end{figure*}

\begin{figure*}
\gridline{\fig{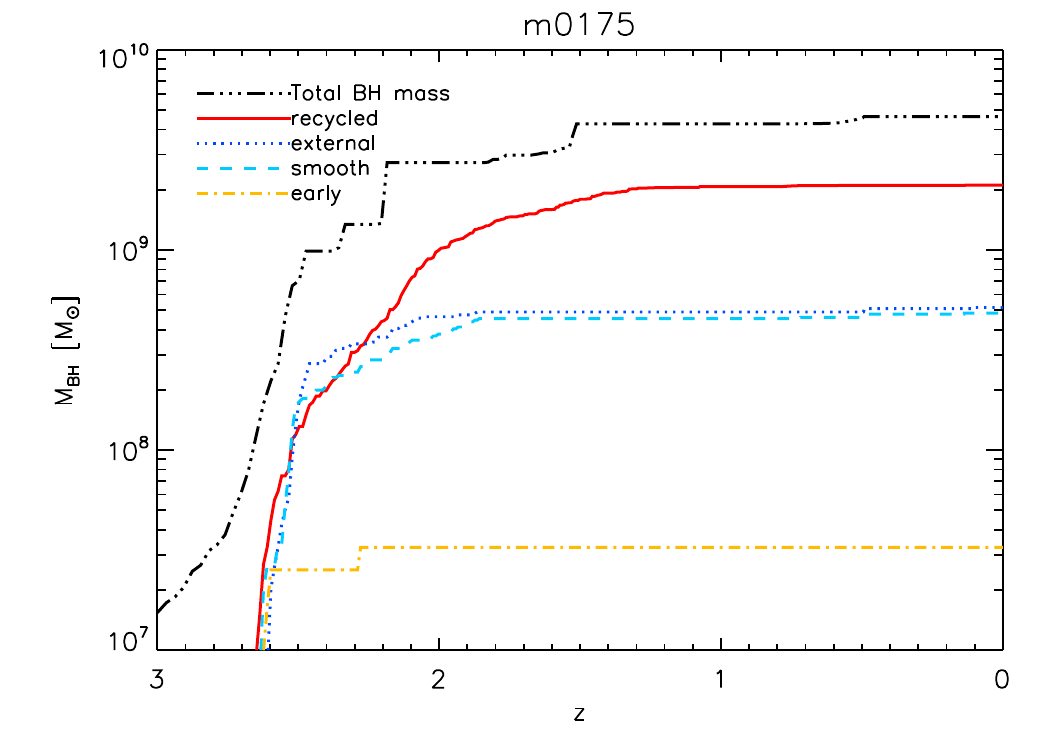}{0.5\textwidth}{}
     \fig{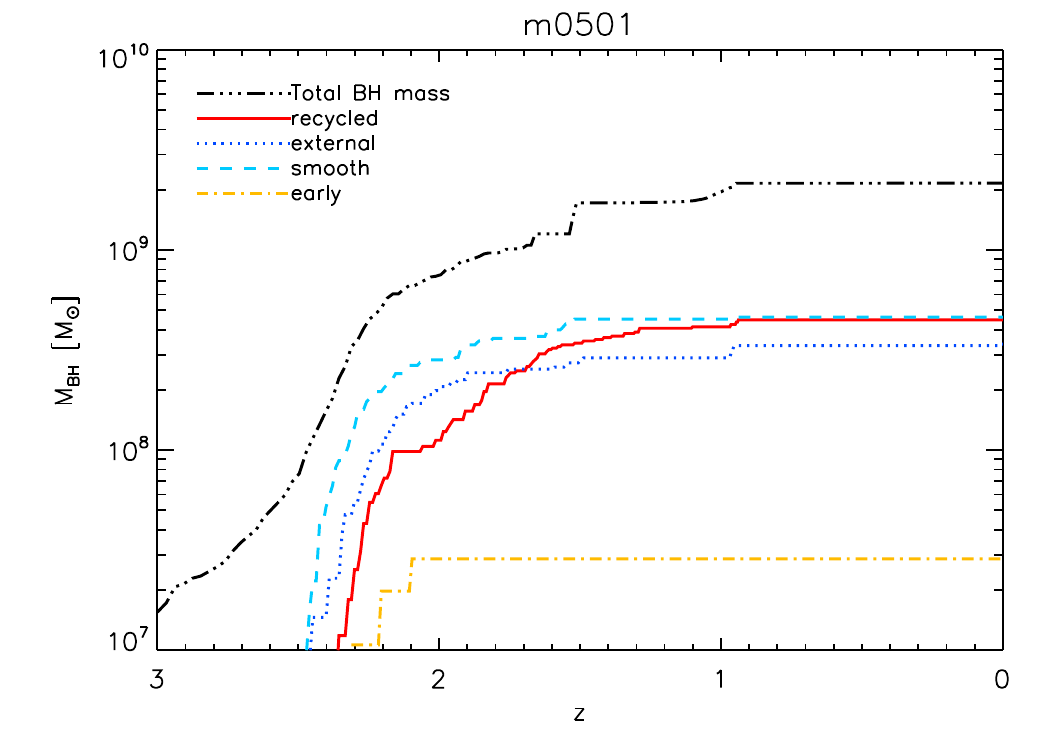}{0.5\textwidth}{}}
\caption{Mass growth of the primary black hole in two example galaxies, m0175 and m0501. The solid black line shows the growth of the total mass of the black hole. The contribution of accreted gas originating from recycled gas, external gas, smoothly accreted gas, and early accretion are shown in red, blue, light blue, and orange, respectively. \label{fig:bhgrowth}} 
\end{figure*}

\begin{figure*}
\gridline{\fig{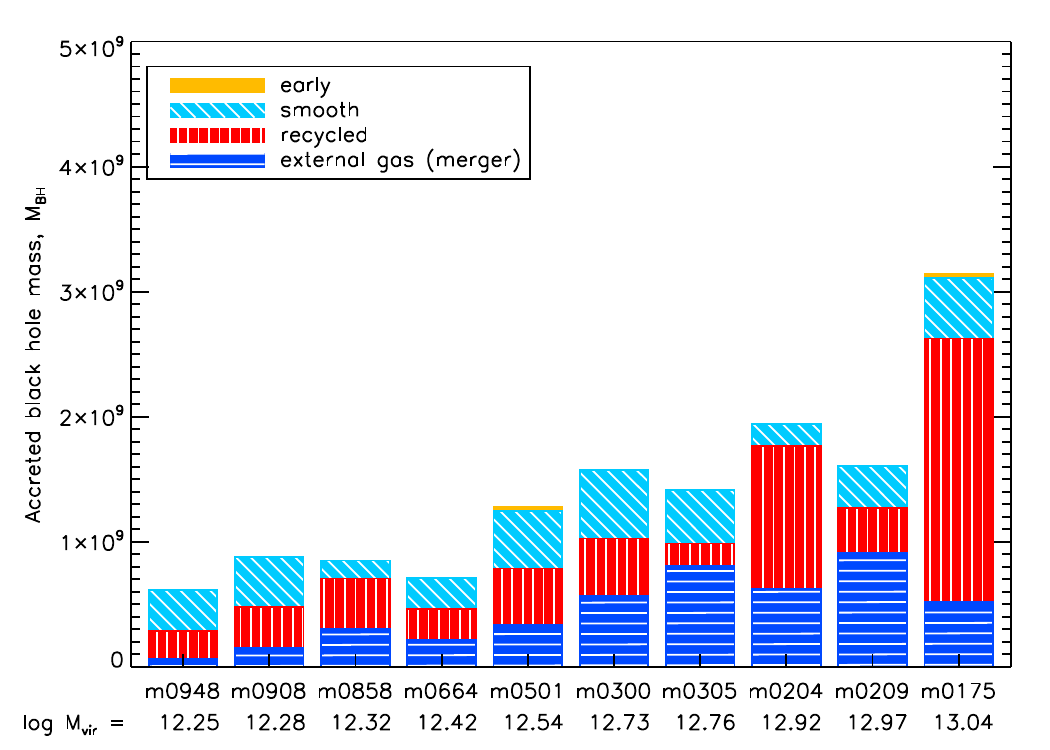}{0.5\textwidth}{}
     \fig{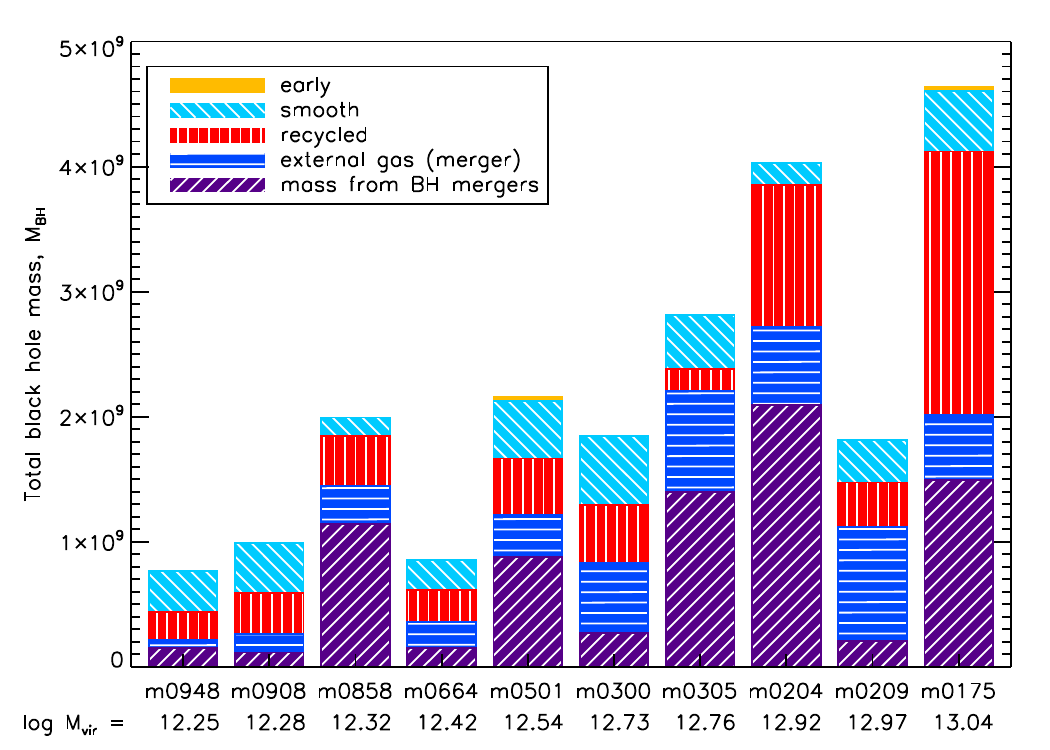}{0.5\textwidth}{}}
\caption{Census of the origin of gas that has been accreted to central black holes as ``early'' (orange), ``smooth'' (light blue), ``recycled'' (red), and ``external'' (blue) per each simulated galaxy. The total virial masses of each of the ten galaxies are listed below. Recycled gas is the primary fuel source for most black holes. In the right panel, we also show mass contribution by black hole mergers (purple). In most SMBHs, the contribution of black hole mergers to the final SMBH mass is smaller than that of gas accretion. \label{fig:censusBH}} 
\end{figure*}

\subsection{Tracing Gas Accretion} \label{subsec:trace}
Particle tracing is a simple yet effective method for investigating the origin of galaxy structures \citep{Oser2010} or baryonic gas cycles \citep{Angles-Alcazar2017c}. In this study, we employed particle tracing to determine the origin of the gas particles accreted by the central black holes. We first compiled a list of gas particle IDs and the IDs of black hole particles to which the gas is accreted whenever gas accretion occurred during the simulations. Then, upon completion of the simulations, we traced the history of each gas particle from the moment it was accreted onto the black hole back to its birth.

We classify the origin of accreted gas particles based on how they enter the primary halo. (1) Gas particles that ever belonged to a different galaxy before accretion are defined as {\bf external}, and (2) all smoothly accreted gas particles (which were never within another galaxy) are labeled as {\bf smooth}. (3) If the gas mass is produced by stellar evolution via AGB winds or SN winds within the primary halo, we classify them as {\bf recycled}. Finally, (4) if gas particles are already in the main halo before $z=5$, they are defined as {\bf early}. 

In \autoref{fig:flowchart}, we show the flowchart of our classification method. For each gas particle accreted by a central black hole, we trace its history to examine its origin. We first label the gas particle {\bf early} if it was in the central halo at $z=5$ or swallowed by a black hole before $z=5$. Next, we check if the gas particle has ever belonged to other halos or subhalos, and label it {\bf external} if it was ever within the virial radius of halos or subhalos other than the central halo. These external gas particles usually enter the central halo via galaxy mergers. Then we check if the gas particles were spawned within 10 \% of the virial radius of the primary halo due to the gas mass ejected from dying stars and label them {\bf recycled}. Additionally, we classify as recycled gas the `mass increment' of gas particles occuring within 10 \% of the virial radius due to stellar ejecta. {\bf Our definition of recycled gas encompasses solely the newly created gas mass inside the galaxy. Specifically, only the fraction of the gas particle that increased mass due to stellar ejecta is classified as 'recycled'.} This recycled gas is usually metal-enriched as winds continuously add it from young stars, supernovas, and AGB stars in our simulation. Finally, all other gas particles that smoothly enter the central halo are labeled {\bf smooth}.  


Note that we only trace the gas particles eaten directly by black holes within the central primary halo. All gas accretion that happened outside of $r_{10}=0.1 \times \rvir$ of the primary halo, e.g., swallowed by black holes in merging galaxies or satellite galaxies, are not included in our analysis. We separately discuss the black hole mass budget contributed by black hole mergers in \autoref{sec:bhmerger}. %

In our analysis, we only count the recycled gas generated within the $r_{10}=0.1 \times \rvir$ of the primary halo as `recycled'. All gas produced from dying stars in external galaxies and then accreted into the main halo later are defined as `external'. %

\section{Results} \label{sec:results}
\begin{figure}
\plotone{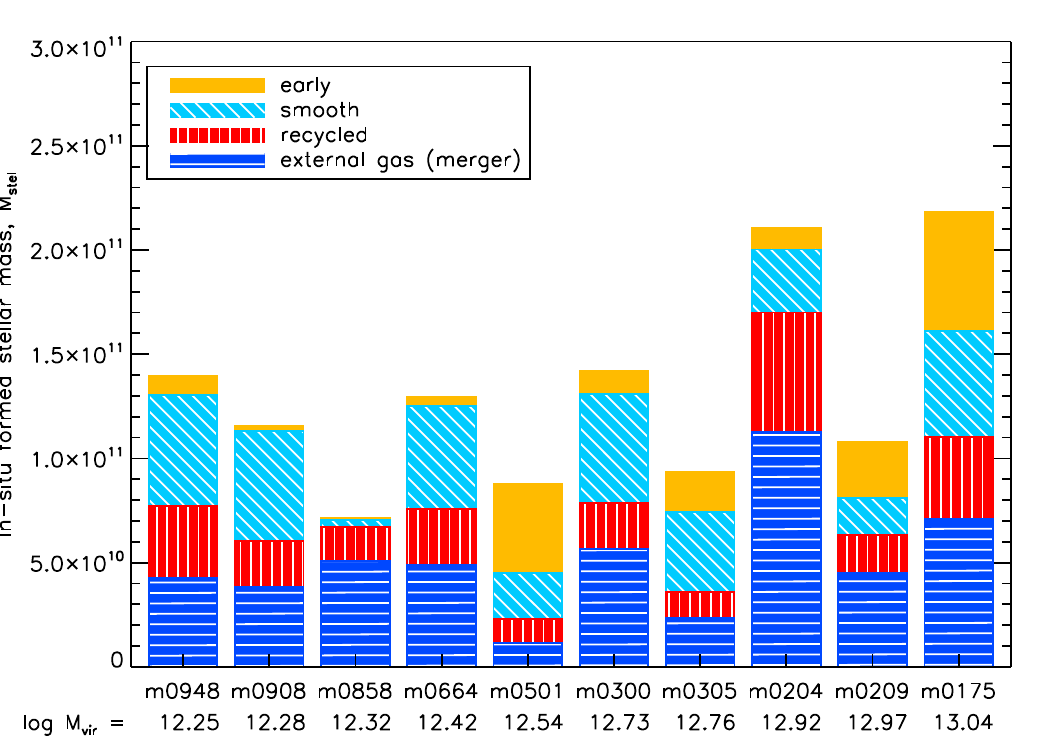} 
\caption{Census of the origin of gas that has been turned into stars within the 10 \% of virial radius of the host galaxy. The fraction of origin of in-situ star forming gas are shown as ``early'' (orange), ``smooth'' (light blue), ``recycled'' (red), and ``external'' (blue) per each simulated galaxy. The total virial masses of each of the ten galaxies are listed below. The external gas is the primary fuel for star formation in many galaxy cases. \label{fig:massfractionSF}} 
\end{figure}

\begin{figure*}
\epsscale{.7}
\plotone{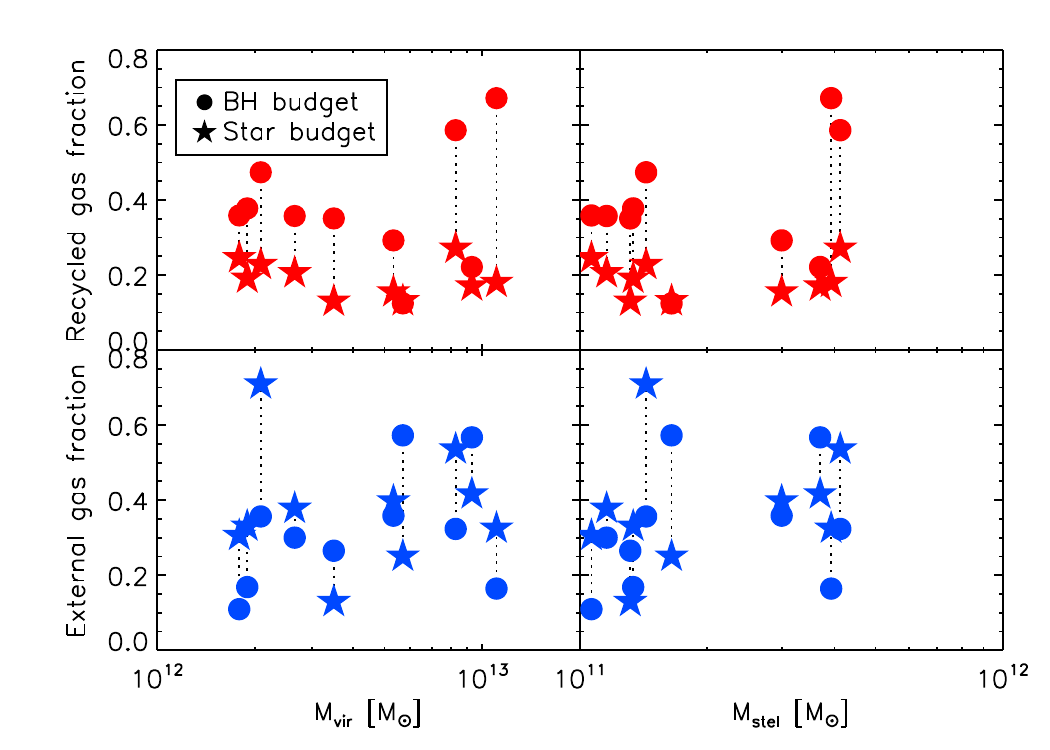}
\caption{Mass fraction of the recycled gas (top) and external gas (bottom) as a function virial mass (left panels) and stellar mass (right panels) for black hole mass accretion (circles) and in-situ star formation (stars). The fractions of the black hole mass budget and star formation budget for the same galaxy are connected by a dotted line. The fraction of recycled gas in black hole gas accretion is {\it always} higher than that in star formation, indicating that recycled gas is a more efficient fuel source for black holes. \label{fig:massfraction_compare}}
\end{figure*}

First, we show the results of our gas history tracing. \autoref{fig:trajectory} shows the trajectory of the BH-accreted gas particles before they are accreted onto the central black holes in two example galaxies. On the y-axis, we show the distance of each gas particle from the central black hole in units of the virial radius of the main halo as a function of redshift. Therefore each line represents the trajectory of a single gas particle before it reaches the central black hole sitting at the bottom of the plot. Most of the gas accretion occurs at $3<z<1$ in the simulated galaxies. Many external gas particles orbit around the galaxy before being accreted. In contrast, recycled gas is produced within the galaxy and almost immediately accreted onto the black hole.

\autoref{fig:bhgrowth} shows the black hole growth over time and how each growth channel contributes to the mass growth in two example galaxies. The black line on the top shows the total black hole mass, including the mass growth through black hole-black hole mergers. With each colored line below, we show the cumulative mass accretion contribution by each gas origin. These black holes stay dormant most of the time, not accreting much gas, especially at late times. This is expected from the observation that most galaxies in the local universe do not exhibit significant AGN activity, but in a small fraction of all galaxies cold gas accreted onto black holes triggers AGN with powerful outflows and intense radiation. For the black hole in the halo m0175 shown, the most massive in our sample, recycled gas is the dominant source of black hole fuel, especially at late times. In the case of m0501, accretion by smooth gas dominates at all times, but the contribution of recycled gas increases at late times, resulting in a comparable contribution by $z=0$.

How much mass is contributed by each type of gas accretion? To investigate the mass budget of the central SMBH at $z=0$, we examine the contribution of each black hole accretion origin. The left panel of \autoref{fig:censusBH} presents a bar plot summarizing the contributions of different black hole accretion origins for all ten SMBHs in our sample. Recycled gas is the primary fuel source for most black holes, such as those in galaxies m0175 and m0204. However, in some cases, external accretion is more dominant than recycled gas accretion, as in galaxies m0209 and m0305. While the contribution by smoothly accreted gas is relatively minor for most SMBHs, it contributes a higher fraction in SMBHs in lower mass galaxies, such as those in m0908 and m0948. 

It has been suggested that in gas-rich galaxy mergers, AGN can be triggered as gas loses angular momentum and flows inward towards the galaxy center and onto the black hole. This merger-driven AGN activity has been shown to occur in idealized simulations of binary mergers of two gas-rich disk galaxies \citep[e.g.][]{2005MNRAS.361..776S}. 
However, as shown in \autoref{fig:censusBH}, black hole growth from \emph{external} gas accreted through galaxy mergers is not dominant in our simulations, at around 30 \% on average, when integrated over the entire evolutionary history of SMBH formation. This is because the cosmological simulations used in this study include cosmological infall, a more complex merger history including dry minor mergers, and recycled gas. Further discussion on this topic will be presented in~\autoref{sec:agn-merger}.  

The right panel of \autoref{fig:censusBH} illustrates the contribution of black hole mergers to the total mass budget. The fraction of the mass budget from BH-BH mergers varies considerably depending on the galaxy's (and black hole's) merger history. In most SMBHs in our sample, the contribution of black hole mergers to the final SMBH mass is smaller than that of gas accretion. 

We now investigate whether recycled gas is the most efficient fuel source for black holes or if it is simply the most abundant compared to other gas origins in host galaxies. To do this, we perform a similar gas history tracking analysis, but this time for all stars born within the galaxy. We trace the history of all gas particles that have been converted into stars within 10 percent of the virial radius of the main halo, i.e., all in-situ formed star particles. This allows us to determine the origin of the gas particles that fuelled the formation of in-situ born stars. 

\autoref{fig:massfractionSF} shows the contribution of each gas origin (early, external, smooth, and recycled) to star formation within the galaxy. The bar plot demonstrates that external gas is the primary fuel for star formation in many galaxies. Recycled gas ejected from evolved stars also contributes to new star formation within the galaxy, but its fraction is much smaller than that in black hole mass budget. Instead, in-situ star formation has a higher contribution from early and smooth accretion compared to black hole feeding. Note that this plot exclusively shows the stellar mass produced within 10 \% of the virial radius of the galaxy, i.e., the stellar mass produced through in-situ star formation. The total stellar masses, which encompass accreted mass from merged galaxies, are more massive than in-situ formed stellar mass shown here. The stellar dispersion and the black hole mass of simulated galaxies follows the local M-sigma relation (see \citet{Choi2017}).

In \autoref{fig:massfraction_compare}, we present the fraction of recycled and external gas contributions to both the black hole and star formation mass budgets, as a function of the galaxy's virial and stellar mass. For the black hole mass budget, recycled gas typically accounts for $38\pm16$ percent of the total gas accretion onto black holes, while external gas contributes approximately $32\pm16$ percent. Among the four gas origin categories, recycled gas makes the most substantial contribution to black hole mass growth. For the star formation mass budget, the contribution of recycled gas is approximately $20$ percent. As shown in the top panel of \autoref{fig:massfraction_compare}, the fraction of recycled gas in black hole gas accretion is always higher than that of star formation, indicating that recycled gas is indeed the most efficient fuel source for black holes. 

\begin{figure}
\plotone{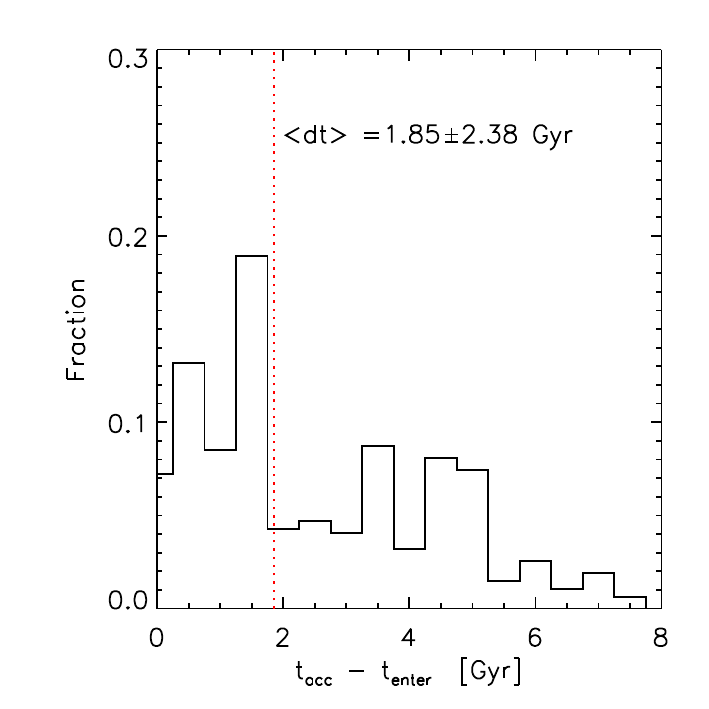}
\caption{The distribution of gas accretion timescales $\Delta t$ for black hole gas accretion with the {\it external} origin. The gas accretion timescale is defined as the time difference between the gas entering the galaxy and the time it actually accretes into the black hole. In other words, it is the time spent by external gas inside the galaxy after entering the virial radius until it finally accretes onto the black hole. \label{fig:dt_bh}}
\end{figure}

On the other hand, as shown in the bottom panel of \autoref{fig:massfraction_compare}, there is no consistent trend in the external gas fraction, unlike the recycled gas fraction. In seven galaxies, the external gas fraction of the star formation budget was higher than that of black hole feeding, while it was lower in the remaining three galaxies. Additionally, neither the external gas fraction nor the recycled gas fraction show a clear correlation with the virial or stellar masses of the galaxy halo. Although higher mass systems are expected to produce more recycled gas and accrete more external gas, no clear correlations were found between these two fractions and galaxy masses. 

Lastly, we investigated the accretion timescale of gas with an external origin. We defined $t_{\rm enter}$ as the moment when a gas particle enters the primary halo, signifying its passage through the $\rvir$ of the main halo. Similarly, $t_{\rm acc}$ represents the moment when the gas particle is accreted by the black hole (BH). Thus, the accretion timescale, denoted as $\Delta t$, for external gas particles is given by $t_{\rm acc} - t_{\rm enter}$. It represents the duration from when the gas particle enters the primary galaxy halo through a merger event to when it ultimately accretes onto the central black hole. We compute $\Delta t$ for all gas particles with external origin.

\autoref{fig:dt_bh} illustrates the distribution of gas accretion timescales $\Delta t$ for all external gas particles. The distribution exhibits a wide range, spanning from a minimum of $\sim 50$ Myr to a maximum of $\sim 7.6$ Gyr, with a mean value of 1.85 Gyr. This indicates that in certain cases, gas from a merger promptly fuels the black hole following the galaxy merger, while in other cases, it resides in the main halo for an extended period before undergoing accretion. This time delay likely contributes to the apparent lack of a clear observed connection between galaxy mergers and AGN activity. Further discussion on this topic will be presented in~\autoref{sec:agn-merger}.

\section{Discussion}\label{sec:discussion}
\subsection{The Importance of Recycled gas}\label{sec:recycled}

Among the four gas origins discussed in this paper -- external, smooth, early, and recycled -- the most significant contributor to the gas accretion of SMBHs is the recycled gas. Although the fraction varies depending on the individual evolutionary history of the SMBH and its host galaxy, recycled gas generated within the host galaxy provides about 40\% of the fuel for AGN on average.

By analyzing the origin of gas that triggers star formation within the galaxy (\autoref{fig:massfractionSF}), we found that the dominant source is external gas. Recycled gas also contributes to star formation, but its fraction is much smaller compared to the black hole mass budget. We found that the recycled gas fraction of the black hole mass budget is always higher than the recycled gas fraction of the star formation mass budget (\autoref{fig:massfraction_compare}). However, it does not indicate that a larger amount of the total recycled gas emitted by evolved stars feeds black holes. In fact, since the total amount of gas consumed by star formation is much greater than the total amount of gas accreted by black holes, the absolute amount of recycled gas consumed by star formation is much greater than the amount accreted by black holes. 

The higher recycled gas fraction of the black hole mass budget compared to the star formation mass budget indicates that the recycled gas feeds black holes the most efficiently among the four origins of gas entering the host galaxy. First, recycled gas has a short cooling time because of its high metallicity so quickly cools down and falls to the center of the galaxy. Second, recycled gas is abundantly produced near the center of the galaxy, around the black hole. In the latter part of the simulation, host galaxies form high-mass elliptical galaxies, and old stellar populations accumulate abundantly in the center. Approximately 30~\% of the initial mass of star particles is released as recycled gas; therefore a significant amount of recycled gas is expected to be released in the vicinity of black holes. 

The importance of recycled gas for fueling ongoing star formation has been discussed in many previous studies using simulations \citep[e.g.][]{Leitner2011}. Recent cosmological hydrodynamics simulations have revealed that recycled stellar mass loss can serve as a trigger for late-time star formation \citep[e.g.][]{Segers2016}. \citet{Ciotti2007} demonstrated recycled gas can trigger star formation and also activate AGN in a hydrodynamic simulations of an isolated elliptical galaxy. However, this study did not involve cosmological infall or mergers. Not surprisingly, recycled gas was the only mechanism that was able to trigger AGN late flaring in an isolated elliptical set-up with a plenty of old stellar population. Our study, on the other hand, employs cosmological simulations, which include both hierarchical galaxy mergers and cosmological infall. We show that even in this simulation setup, recycled gas still contributes significantly to black hole growth.

Observationally, there are many examples of AGN that are believed to have been triggered by a contribution from recycled gas. Recently, various studies have shown that the AGN fraction in quiescent galaxies is quite high, and it has been speculated that it may be fueled by stellar winds \citep{Aird2018,Aird2019,Birchall2023}. In particular, the AGN fraction of the most recently quenched extended quiescent galaxies, expected to have high stellar mass loss, was surprisingly high as 10-30 \% \citep{Aird2022}. There also have also been a number of recent observational studies of the relationship between the age of the stellar population in host galaxies and AGN activity. \cite{Riffel2022,Riffel2013} found a correlation of the intermediate stellar population with AGN luminosity, which is interpreted as the recycled material from intermediate age stellar population is triggering the AGN. \cite{Ni2023}  recently showed that a higher lever of AGN activity is observed in galaxies with younger stellar population ages, where more recycled gas is produced. Consistent with these observational findings, our study highlights the substantial contribution of recycled gas in triggering AGN within a comprehensive cosmological hydrodynamic simulation.

\subsection{AGN-merger connection}\label{sec:agn-merger}
The hypothesis proposing that galaxy mergers trigger AGN has been extensively examined \citep[e.g.][]{Cisternas2011,Kocevski2015,Bickley2023,Dougherty2023}. Notably, simulations of isolated gas-rich disk mergers have demonstrated the activation of a strong AGN \citep{2005MNRAS.361..776S,2008ApJS..175..356H}. Numerous observational studies have sought to explore the connection between AGN and mergers by comparing the merger fraction of AGN-hosting galaxies to that of matched samples of non-AGN galaxies. However, these investigations have generated conflicting results. Some studies have shown that gas-rich mergers are associated with enhanced BH activity \citep{2011MNRAS.418.2043E,2011ApJ...743....2S,Ellison2019a}. However, some studies such as \citet{Kocevski2012,Villforth2014} have found that AGN do not exhibit a notably higher merger fraction when compared to a control sample of galaxies with similar properties. {There are many examples of significant AGN activity, such as AGN winds, being observed in galaxies that do not have merger signatures \citep[e.g.][]{Smethurst2021}.}

Our study provides compelling insights into the underlying reasons why the AGN-merger connection may not be seen in observational studies. Note again {\it mergers do fuel AGN in our simulations}. Our simulations showed that about 30\% of the total gas consumed by the BH have external origin, and entered the primary halo through mergers and subsequently accreted onto the BH. 

However, the lack of an observational detection of a higher merger fraction in AGN host galaxies relative to the control sample may be attributed to two primary reasons:

\begin{enumerate}
\item The quantity of gas that directly enters through a merger and triggers an AGN is relatively small. As illustrated in \autoref{fig:censusBH}, the fraction of externally sourced gas, on average, constitutes around 30\% of the total gas accreted by the BH. 

\item Furthermore, there is often a considerable time delay between a galaxy merger and the actual accretion of external gas onto the BH. As indicated in \autoref{fig:dt_bh}, the average time delay is approximately 1.85 Gyr. The gas entering the virial radius of the primary galaxy via a merger takes a considerable amount of time to shed its angular momentum before it can accrete onto the central black hole. This time lag between the appearance of a merger and the actual triggering of the AGN can result in a diminished merger fraction among AGNs.
\end{enumerate}

Hence, the presence of a paired or closely interacting companion galaxy does not necessarily indicate the immediate triggering of an AGN, as the gas entering the primary halo through the merger process may not have initiated AGN activity {\it yet}. Therefore, identifying mergers solely based on companion galaxies (double nuclei, pairs, etc.) may be insufficient for assessing the connection between AGN and mergers. Instead, it is more effective to trace post-merger signatures, such as tidal structures that leave morphological imprints over an extended time period, as a means of identifying AGN-merger connection. Indeed, many observations have also revealed that most AGN are in the post-coalescence phase of mergers, showing disturbed morphologies \citep[e.g.][]{Almeida2011,Almeida2012,Bessiere2012}.

However, here we only investigate the role of gas that enters the main galaxy from an external galaxy via a merger. It is essential to note that this study does not investigate the role of BH fueling by gas that is pre-existing within the main galaxy, and driven into the nucleus by a merger-triggered gravitational instability. Mergers inherently redistribute angular momentum from the central region of the galaxy to its outer regions, causing mass to move inwards. Consequently, gas that would have otherwise been rotationally supported in the absence of a merger may by driven into the nucleus, ultimately accreting onto the black hole. The accurate tracking of this type of AGN triggering by mergers necessitates an alternative approach, one that we intend to address in future work.

In summary, galaxy mergers undeniably have the potential to trigger AGN. However, it is important to acknowledge that the direct external gas supply for black hole accretion through mergers is constrained within the context of a fully cosmological simulation framework. Our work emphasizes the importance of internal processes, particularly the impact of recycled gas from stellar evolution, in driving AGN activity. \citet{steinborn2018} recently reported a similar finding, demonstrating a minor impact of mergers on a sample of AGN derived from a comprehensive cosmological simulation spanning a large volume. \cite{Martin2018,Smethurst2024} also showed that only 35\% of the cumulative growth of black holes is due to mergers in Horizon-AGN simulations.

For a detailed analysis of the AGN-merger connections in the same simulation suite used for this study, we refer the reader to \citet{Sharma2024}. In this paper, synthetic images resembling Hubble Space Telescope (HST) observations were generated using post-processing radiation transfer techniques. The morphological properties of mock images were examined to demonstrate the subtlety of the AGN\--merger connection.

\subsection{Contribution of black hole mergers}\label{sec:bhmerger}
Given that our study primarily focuses on investigating the origin of the gas responsible for activating the AGN in the primary halo, we conducted a comprehensive analysis of gas accretion history within 0.1 $\rvir$ of the primary halo. It is worth noting that neighboring satellite galaxies may also harbor their own black holes, which accrete gas independently. Consequently, during galaxy mergers, black holes that have grown via their own accretion will eventually coalesce with the main black hole. However, the contribution of these black hole mergers to the overall black hole mass budget is relatively minor.  \autoref{fig:censusBH} illustrates the substantial variation in the BH-BH hole merger contribution, depending on the specific black hole merger history. A more in-depth examination of the black hole merger history is beyond the scope of this paper and will be conducted in a follow-up study.

\subsection{ Comparison with previous studies}
In this subsection, we compare our findings with previous studies employing direct particle tracing methods to investigate the origin of gas accreted into black holes.

Firstly, \cite{Bellovary2013} explored the formation origins of three black holes in the mass range of $10^{6-7} \Msun$, simulated up to redshift $z=4$. These moderate-mass black holes are situated at the cores of galaxies with stellar masses ranging from $10^9$ to $10^{10} \Msun$. Their analysis indicates that these black holes predominantly accreted cold-flow gas rather than gas resulting from mergers. However, it is crucial to note that, compared to the gas constituting galaxies, host galaxies contain a substantial amount of cold-flow gas. Their work suggests that the gas content of black hole accretion is ultimately influenced by the gas content of the host galaxy halo.

Meanwhile, \cite{Sanchez2018} investigated the origin of a $\sim 10^7 \Msun$ black hole, simulated down to redshift z=0. This black hole resides at the center of a Milky Way-sized galaxy characterized by a formation history rich in mergers. Reflecting the formation history of the host galaxy, the primary fraction of gas accreted by the black hole was of 'clumpy' origin, i.e., gas with an external source as defined in our study. Their Figure 2 indicates that 56\% of the gas comprising the galactic halo had a clumpy (external) origin, whereas the fraction accreted by the black hole was 74\% clumpy (external). Notably, the fraction of clumpy-origin gas significantly increases toward the galaxy's center. They show that galaxies entering the galactic halo through mergers have lower angular momentum, making them more prone to central accumulation.

In our study, we examined ten black holes with masses of $10^{8.9-9.7} \Msun$ in more massive galaxies. Due to the nature of these host galaxies, there is relatively little gas that ends up within the galaxy, inside 0.1 $\rvir$. Consequently, we analyzed the origin of gas that underwent star formation within this region and compared the fraction of gas feeding the black hole with the fraction feeding the galaxy in each mode. In contrast to the findings of Sanchez et al., where the external (clumpy) fraction of the black hole was larger than that of the galactic halo, our study reveals variability in these external gas fractions depending on the specific galaxy case (refer to \autoref{fig:massfraction_compare}).

Furthermore, our study delves into the analysis of the fraction of recycled gas, a facet that has not been explored through particle tracing in prior studies. The significance of recycled gas becomes more apparent in our study, given that the galaxies examined are more massive compared to those in earlier investigations. 

\subsection{Caveats and limitations}
Observations of AGN have shown that the accretion flows around black holes are composed of cold and hot gas. These accretion flows are theoretically described by several models, such as the standard thin disk \citep{Shakura1973} and the advection dominated accretion flow \citep[e.g.][]{Rees1982} which are on subparsec scales. However, cosmological simulations of galaxy formation do not resolve the black hole accretion disk scale, as the vast dynamic range of cosmological simulations makes this infeasible. With limited spatial and mass resolution, many contemporary simulations adopt a simple black hole accretion prescription, such as a Bondi parameterization \citep{1944MNRAS.104..273B,1952MNRAS.112..195B}. This means that cosmological galaxy formation simulations do not resolve gas transport down to the inner accretion disk surrounding the central massive black hole. 

The simulations used in this study also adopt a Bondi-Hoyle-Lyttleton parameterization \citep{1939PCPS...34..405H,1944MNRAS.104..273B,1952MNRAS.112..195B}. As discussed in \autoref{subsec:simulation} and in \citet{Choi2015a} more in detail, we additionally implemented other sub-grid aspects of the accretion model: a soft Bondi criterion that statistically limits the accretion to the gas within the Bondi radius and the free-fall timescale that accounts for the time it takes to be accreted onto the black hole. Still, we do not fully resolve gas transport down to the inner subparsec.

The cosmological simulation used in this study does not resolve black hole accretion on subparsec scales, but it is suitable for fully tracing the origin of the cosmological gas inflows that ultimately end up near the black hole at the center of the galaxy. Physical processes occurring at scales below the simulation resolution limit are not expected to significantly impact the type of gas that fuels the black hole. However, follow-up research will be necessary to investigate how the gas accretion budget differs in other accretion models, such as the gravitational torque method \citep{Angles-Alcazar2017a}, or at higher resolutions \citep[e.g.][]{Angles-Alcazar2021}.

In this simulation, a simple black hole seeding model is used, where a black hole seed of $10^5 \thinspace \Msun$ is generated at the center of each newly formed dark matter halo above a certain mass limit. However, such a small seed mass is unlikely to have a significant impact on the growth of a supermassive black hole with a mass of $\sim 10^9 \thinspace \Msun$. The specific details of early black hole growth may vary depending on the adopted black hole seeding prescriptions, mimicking various black hole seeding theories, such as Pop III, nuclear star cluster, and direct collapse black seed formation. Nevertheless, given the current resolutions, black hole accretion before $z > 4$ is quite negligible (See~\autoref{fig:trajectory}). The effect of various seeding models on early black hole growth is outside the scope of this study and will require new simulations focused on the early growth of black holes at higher resolution.

Finally, we did not include tidal disruption events (TDEs), which occur when stars approach close to black holes and are torn apart by the tidal force of the black hole. However, the contribution of the accretion of stellar material onto the final SMBH is expected to be negligible, as the observed TDE rate is estimated to be approximately $\sim 10^{-5} \rm yr^{-1}$ per galaxy \citep[e.g.][]{VanVelzen2014,Komossa2015}. This rate is significantly lower than the average gas accretion rate.

\section{Summary}\label{sec:summary}
In this study, we employ particle tracing techniques to investigate the accretion history of all gas particles consumed by $\sim 10^9$ solar mass black holes (SMBH) in a suite of hydrodynamic zoom-in simulations. Based on the origin of each gas particle, we classify them into four categories: external gas originating from other galaxies, recycled gas produced from stellar losses, smoothly accreted gas that quietly enters the main halo, and early gas that was present in the main halo since early times. By examining the formation of ten massive halos and the origins of SMBH accretion within the centers of the main galaxy they host, we draw the following conclusions.

\begin{enumerate}
\item Recycled gas is the dominant source contributing to SMBH accretion, accounting for approximately 38 \% of the total accretion on average. In the massive evolved galaxy systems in our simulations, the influence of the old stellar population is especially significant.
\item The second most substantial contribution comes from an external origin. In cosmological hydrodynamic simulations, galaxies experience diverse hierarchical merger processes. On average, about 32 \% of the gas that enters a main galaxy through mergers, having previously belonged to another galaxy, subsequently accretes onto a black hole.
\item Depending on the galaxy and black hole's formation and merger history, there are instances where the external fraction surpasses the recycled fraction. However, in eight out of the ten  cases we studied, the external fraction does not dominate the total accretion.
\item The time delay between the entry of external gas into the galaxy and the onset of AGN activity exhibits a broad distribution, with an average delay of 1.85 Gyr. This significant time delay, coupled with the finding that external gas is not the primary contributor to gas accretion onto the black hole, may provide a partial explanation for why AGN do not always appear to be associated with mergers.

\end{enumerate}

\acknowledgments
We express our gratitude to the anonymous referee for providing many valuable comments and suggestions. We thank Sara Ellison, Cristina Ramos Almeida, Rogério Riffel, Jong~Hak Woo, and Sukyoung Yi for helpful conversations and suggestions.  This work was supported by the National Research Foundation of Korea grant funded by the Korea government (MSIT) (No. RS-2023-00213322). R. Somerville gratefully acknowledges support from the Simons Foundation. M. Hirschmann acknowledges funding from the Swiss National Science Foundation (SNF) via a PRIMA grant PR00P2 193577 `From cosmic dawn to high noon: the role of BHs for young galaxies’. T. Naab acknowledges support from the Deutsche Forschungsgemeinschaft (DFG, German Research Foundation) under Germany’s Excellence Strategy - EXC-2094 - 390783311 from the DFG Cluster of Excellence ``ORIGINS''. 

\bibliography{library}

\end{document}